\documentclass[journal]{IEEEtran}

\usepackage{graphicx}%
\usepackage{multirow}%
\usepackage{amsmath,amssymb,amsfonts}%
\usepackage{amsthm}%
\usepackage{mathrsfs}%
\usepackage[table]{xcolor}
\usepackage{xcolor, xurl}%
\usepackage{textcomp}%
\usepackage{manyfoot, unicode}%
\usepackage{booktabs}%
\usepackage[english]{babel}
\usepackage[utf8]{inputenc}
\usepackage{nicematrix}
\usepackage{subfig}
\usepackage{caption}
\usepackage[fontsize=11pt]{fontsize}
\usepackage{hyperref}
\usepackage{academicons}
\usepackage[column=0]{cellspace}

\usepackage{etoolbox}
\patchcmd{\appendix}{\appendixname}{\appendixname\ }{}{}

\newcommand{\etal}{et al.}
\begin{document}

\title{Enhancing Prostate Cancer Diagnosis with Deep Learning: A Study using mpMRI Segmentation and Classification}

% \author*[1]{\fnm{Anil B.} \sur{Gavade}}
% % \email{abgavade@git.edu}
% \author*[2]{\fnm{Neel} \sur{Kanwal}}
% % \email{iiauthor@gmail.com}
% % \equalcont{These authors contributed equally to this work.}
% \author[3]{\fnm{Priyanka A.} \sur{Gavade}}
% \author[4]{\fnm{Rajendra} \sur{Nerli}}
% \affil[1]{\orgdiv{ Department of E\&C}, \orgname{KLS Gogte Institute of Technology}, \orgaddress{\city{Belagavi}, \postcode{590008}, \country{India}}}
% \affil[2]{\orgdiv{ Department of Electrical Engineering and Computer Science}, \orgname{University of Stavanger}, \orgaddress{\city{Stavanger}, \postcode{4036}, \country{Norway}}}
% \affil[3]{\orgdiv{ Department of Computer Science and Engineering}, \orgname{Dr. M.S. Sheshgiri College of Engineering and Technology}, \orgaddress{\city{Belagavi}, \postcode{590008}, \country{India}}}
% \affil[4]{\orgdiv{ Department of Urology}, \orgname{JN Medical College}, \orgaddress{\city{Belagavi}, \postcode{590008}, \country{India}}\\
% * These authors contributed equally to this work.\\ 
% Corresponding author: abgavade@git.edu\vspace{-4em}}

% For Arxiv
\author{Anil B. Gavade$^1$*, Neel Kanwal$^2$*, Priyanka A. Gavade$^3$, Rajendra Nerli$^4$\\
$^1$ Department of E\&C KLS Gogte Institute of Technology, Belagavi, India \\
$^2$ Department of Electrical Engineering and Computer Science, University of Stavanger, Stavanger, Norway\\
$^3$ Department of Computer Science and Engineering, Dr. M.S. Sheshgiri College of Engineering and Technology, Belagavi, India\\
$4$ Department of Urology, JN Medical College, Belagavi, India\\
* These authors contributed equally to this work.\\ 
(Corresponding author: abgavade@git.edu)}

%%==================================%%
%% sample for abstract %%
%%==================================%%
\maketitle
\begin{abstract}
Prostate cancer (PCa) is a severe disease among men globally. It is important to identify PCa early and do a precise diagnosis for effective treatment. For PCa diagnosis, Multi-parametric magnetic resonance imaging (mpMRI) emerged as an invaluable imaging modality that offers a precise anatomical view of the prostate gland and its tissue structure. Deep learning (DL) models can enhance existing clinical systems and improve patient care by locating regions of interest for physicians. Recently, DL techniques have been employed to develop a pipeline for segmenting and classifying different cancer types. These studies show that DL can be used to increase diagnostic precision and give objective results without variability. This work uses well-known DL models for the classification and segmentation of mpMRI images to detect PCa. Our implementation involves four pipelines; Semantic DeepSegNet with ResNet50, DeepSegNet with recurrent neural network (RNN), U-Net with RNN, and U-Net with a long short-term memory (LSTM). Each segmentation model is paired with a different classifier to evaluate the performance using different metrics. The results of our experiments show that the pipeline that uses the combination of U-Net and the LSTM model outperforms all other combinations, excelling in both segmentation and classification tasks.
\end{abstract}

% \keywords{Classification and Segmentation, Deep Learning,  Prostate Cancer}

\section{Introduction}\label{sec:intro}
Cancer is the deadliest disease that affects individuals and the global economy. Prostate cancer (PCa) specifically impacts the prostate gland, situated near the rectum and below the bladder~\cite{stats,liu2009prostate}. PCa is one of the most commonly diagnosed malignancies in males and ranks as the fourth most common cancer. Over the years, the number of cases has been steadily increasing, with an estimated 1.4 million patients diagnosed globally in 2020 globally~\cite{stats}. PCa is widely observed in the elderly male population, with around 6 in 10 cases detected in men aged 65 or older, particularly among non-Hispanic black men~\cite{cancer_survival}. If PCa is not detected in its early stages, the fatality rate for PCa can be substantial, with 1 in 41 men losing their lives~\cite{stats,cancer_survival}. Clinically, PCa is detected using different clinical procedures such as, the pathology of prostate tissue, mpMRI scans, and prostate-specific antigen testing. The pathologists, radiologists, and clinicians who perform these procedures invest sufficient time to accurately identify cancer concentrations~\cite{abraham2019automated,gavade2023cancer, tabatabaei2023self}. The treatment of cancer depends on the concentration and severity of the disease. However, manual procedures involving human interventions for PCa detection are time-consuming, often resulting in incorrect and delayed diagnoses.

\begin{figure*}[ht!]%
\centering
\includegraphics[width=1\textwidth]{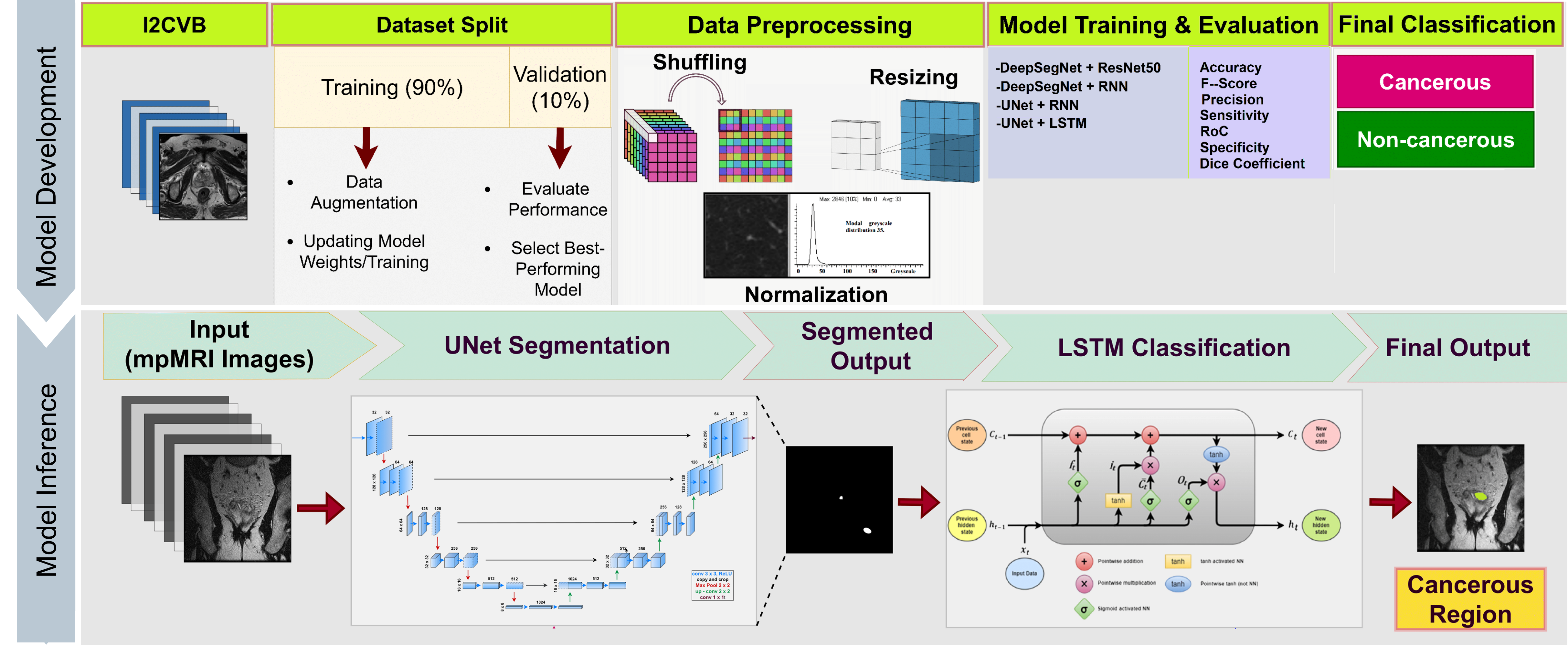}
\caption{\textbf{Illustration of our proposed approach.} \emph{Development:} First step for developing the models that are tested on the validation set. \emph{Inference:} Best models from the first step are deployed for the segmentation of candidate regions and then the classification of those regions.}\label{fig1}
\end{figure*}

With the advancement of neural network applications, deep learning (DL) has evolved, and its performance is of great significance in assessment and is widely employed in biomedical imagery applications. DL algorithms have the great potential to understand intricate patterns and unseen representations from images~\cite{artan2010prostate, gavade_peripheral}. Research in the bioinformatics field of biomedical imagery has garnered attention for discovering the application of DL algorithms in cancer diagnosis ~\cite{kanwal2023detection,gavade2023automated,kanwal2023balancing}. DL algorithms are powerful tools for the segmentation, detection, and classification of PCa, as well as for estimating postoperative reappearance and prognostic outcomes~\cite{gavade_peripheral}.

Non-invasive PCa diagnosis, notably magnetic resonance imaging (MRI), has gained prominence. MRI employs a potent magnetic field, radio-frequency pulses, and advanced processing to produce intricate images, focusing on assessing PCa extent and guiding biopsies. mpMRI includes dynamic contrast-enhanced (DCE), T2-weighted (T2-W), diffusion-weighted imaging (DWI), and magnetic resonance spectroscopy, enhancing the accuracy of detection. In contrast, Bi-parametric MRI (bpMRI) streamlines the process by focusing on T2-weighted and diffusion-weighted sequences. While both offer benefits, mpMRI's comprehensive approach is widely preferred for accurate diagnosis and treatment planning, making it a cornerstone in PCa follow-ups.

mpMRI is a comprehensive imaging technique that offers detailed information about the size, location, and progression of tumors within the prostate gland and surrounding tissues. The mpMRI offers superior imaging of the prostate gland, making it a valuable data source for developing DL models. DL algorithms can accurately detect cancer by identifying specific features in mpMRI images associated with PCa, such as tumor size and location. Computer-aided (CAD) tools using DL methods can provide decision support to clinicians, helping them focus on regions of interest (ROIs) and optimizing time efficiency in the diagnostic procedure~\cite{kanwal2023detection,kanwal2023vision}. CAD tools also provide personalized treatment strategies based on the grade of cancer.  Deep neural networks (DNNs) can readily detect malignancies related to PCa, such as cancerous areas and dimensions, and accurately classify cancer~\cite{mehta2021computer,gavade_peripheral}.  Generalizing DL algorithms for PCa is complex, as it requires access to a sufficient amount of clinical data with accurate labels~\cite{kanwal2023detection}. To compensate for this, we have applied the transfer learning (TL) technique. TL aims to use the knowledge of a model training on a different task to retrain it on a new task, using a new dataset~\cite{kanwal2023detection}.This research paper proposes the segmentation and classification of PCa with the aid of DL, as shown in Figure~\ref{fig1}. 

The proposed pipeline combines DL models in two phases: U-Net/DeepSegNet segmentation architecture and ResNet-50, RNN, and LSTM models for classification. These DL models leverage recurrent networks and convolutional networks to comprehend complex patterns and spatial requirements, resulting in promising outcomes and refining automated diagnosis. The training approach utilizes data augmentation in conjunction with TL methods to overcome the limitations posed by the scarcity of labeled data, resulting in improved precision. All DL models are trained by combining different architectures and are compared with related works(discussed later in section~\ref{sec:related}).
Our paper introduces an innovative DL approach by seamlessly integrating segmentation and classification in an end-to-end fashion.
The proposed DL method outperforms others in terms of effective segmentation and classification of PCawith a promise to revolutionize clinical practice by enabling more precise diagnosis and treatment planning, eventually enhancing
patient care.

\section{Related Work}\label{sec:related}
The application of DL approaches in the medical domain has paved the way for the development of predictive systems for automated diagnosis. In this context, mpMRI is a preferred modality due to its non-invasive data collection technique~\cite{mehta2021computer}. For improved patient care, mpMRI and DL combined in automated CAD systems can save doctors' time and lead to rapid PCa diagnosis~\cite{gavade2023automated}. 

Litjens \etal~\cite{litjens2014computer} proposed a two-stage CAD tool for screening MRI, resulting in 69\% sensitivity and 83\% specificity. In the first stage, they performed voxel feature extraction and later used candidate regions to obtain cancer likelihood. Artan \etal~\cite{artan2010prostate} introduced a system for PCa localization on multi-spectral MRI, demonstrating commendable sensitivity and specificity for cancer detection. Similarly, Peng \etal~\cite{peng2013quantitative} developed a CAD system for PCa detection and distinguished cancerous tissue from regular tissue using mpMRI, achieving a remarkable area under the curve (AUC) of 0.85. Brosch \etal~\cite{brosch2018deep} devised a DL approach for boundary detection using neural networks and segmentation of the prostate in MRI images. Their method demonstrated promising results, achieving a mean absolute distance of 1.4 mm. Liu \etal~\cite{liu2009prostate} recommended a technique for PCa segmentation using MRI, achieving a 0.62 dice score for segmenting the cancerous regions. 

Among approaches utilizing DNN, 
% Tian \etal~\cite{tian2018psnet} proposed a PSNet model for prostate segmentation, achieving an average Hausdorff distance of 4.16 mm on the test dataset.  
Aldoj \etal~\cite{aldoj2020automatic} employed a DenseNet-based U-net model to achieve automatic prostate segmentation using MRI images. They utilized support vector machines (SVMs) and conditional random fields (CRFs) to improve segmentation. 
Karimi \etal~\cite{karimi2018prostate} introduced a custom DNN architecture for prostate segmentation, employing a training setup relying on the statistical shape of the data.
Abraham \etal~\cite{abraham2019automated} utilized a VGG16 DNN with an ordinal classifier to evaluate the Gleason score (GS) for PCa grading, achieving accurate grade prediction and reducing the mortality rate due to PCa. 
Cao \etal~\cite{cao2019joint} recommended FocalNet, a DNN for the simultaneous multiclass detection of PCa prediction and aggressiveness of lesions utilizing GS, contributing to the development of automated clinical decision support systems for PCa diagnosis. 

Benefiting from TL, Zhong \etal~\cite{zhong2019deep} developed a model for classifying PCa in mpMRI. Their approach showed promise in improving PCa detection accuracy and efficiency, providing valuable clinical decision support to radiologists.  Mehta \etal~\cite{mehta2021computer} developed a CAD system using clinical features. Their system distinguished malignant cancer from benign tissue with AUC values of 0.79; and 0.85 AUC for distinguishing between low concentration of cancer and high concentration of cancer, respectively. 
Merging textural and morphological analysis, Zhang \etal~\cite{zhang2021new}proposed an innovative method for PCa detection using MRI, achieving accuracy rates of 89.6\%, sensitivity of 87.5\%, and specificity of 90.8\%. This approach shows promise in elevating the accuracy and efficiency of PCa detection, providing valuable clinical decision support for radiologists. 
Recently, Sun \etal~\cite{sun2023deep} created DNN models for evaluating PCa features which are clinically significant in mpMRI images. Gavade \etal~\cite{deepsegnet} introduced DeepSegNet and trained in CT images to perform semantic segmentation. Later, Gavade \etal~\cite{gavade_peripheral} highlighted the promise of data-driven algorithms in PCa analysis through mpMRI. They pinpointed nnU-Net's promising performance in precisely segmenting PCa across the peripheral zone and central gland regions. Interestingly, ResNet-50 emerged as the frontrunner in classification, showcasing remarkable accuracy. These findings underscore the transformative potential of TL for DL methodologies.

By leveraging state-of-the-art DL models and developing new pipelines on mpMRI data, the emerging research can potentially revolutionize PCa diagnosis. Integrating AI-driven analyses with conventional clinical assessments will ultimately lead to improved diagnostics and prognostics for PCa using mpMRI.

\section{Methods and Materials}\label{sec:method}
\subsection{Data Material}
The I2CVB stands for Initiative for Collaborative Computer Vision Benchmarking dataset is a public resource~\cite{lemaitre2015computer}. The dataset includes annotated mpMRI images captured from two scanners with different resolutions: (a) a General Electric (GE) scanner with a resolution of 1.5 Tesla and (b) a Siemens scanner with a resolution of 3.0 Tesla (discussed in section~\ref{sec:intro}) along with apparent diffusion coefficient maps.

\subsection{Preprocessing}
mpMRI preprocessing is essential for enhancing the classification performance of DL models. In order to reduce image artifacts, we have normalized images in the dataset. To train the model, the datasets were divided into two subcategories: a training subset with 90\% and a validation subset with 10\% of the data. Data augmentation is applied to compensate for the small sample size. By rotating, turning, and translating, as well as by adding noise to images, we were able to produce new samples from already existing ones. Bicubic interpolation was used to calculate per-pixel displacements. A Gaussian distribution with a standard deviation of 10 pixels was used to sample these displacement vectors. Through the use of random displacement vectors and a coarse $3\times 3$ grid, smooth deformations were produced. Annotated images in the training dataset were utilized to train the segmentation network using random elastic deformations in order to give the model acquisition invariance and resilience qualities. Ultimately, the dataset included finally scaled down to $256\times 256$.

\subsection{Deep Learning Techniques}
DNNs serve as the foundational architecture, forming the backbone for pioneering DL models like DeepSegNet~\cite{deepsegnet} and U-Net~\cite{ronneberger2015u}, which revolutionized image segmentation through their intricate feature extraction capabilities. In our architecture, these DL-based segmentation methods seamlessly integrate with state-of-the-art classification models, including ResNet-50, RNN, and LSTM. This seamless integration empowers the overall DL pipeline, elevating the precision of mpMRI classification. Our two-stage DL technique is focused on harnessing the capabilities of two DL models, particularly CNNs, and RNNs, to advance the classification of PCa using mpMRI. Our first stage applies segmentation to find the candidate region. We are using U-Net~\cite{ronneberger2015u} and DeepSegNet~\cite{deepsegnet} for this purpose. In the second stage, segmented candidate regions are supplied for classification. Here we use CNN ResNet-50, RNNs and LSTM. Overall we establish four combinations and train these pipelines on the same training and validation set. We integrate DeepSegNet~\cite{deepsegnet} with ResNet50 and RNN to test how well this combination performs on mpMRI data. Identically, we combine U-Net with RNN and LSTM. An overview of all four pipelines is given in Figure~\ref{fig1}.

\subsubsection{Segmentation}
\vspace{1em}
\textbf{\newline DeepSegNet: } DeepSegNet, originally proposed by Gavade \etal~\cite{deepsegnet}, designed for tasks like a brain tumor and liver lesion analysis, stands as a formidable solution for precise cancer region segmentation. DeepSegNet model consists of an encoder and decoder model, In which the encoder consists of several convolutional layers, which are responsible for learning the features of the input image; while the decoder is composed of several upsampling layers, which increase the spatial resolution of the feature maps and generate the final segmentation mask. DeepSegNet also uses skip connections, which allow the decoder to access features from the encoder at multiple resolutions, improving the accuracy of the segmentation results. 
Furthermore, DeepSegNet inherently encompasses contextual awareness, enabling it to consider both local and global contexts unique features for accurate prostate gland segmentation within the larger anatomical landscape.
%% ---- ------
%\newline
\newline
\textbf{U-Net: } U-Net,developed by Ronneberger \etal~\cite{ronneberger2015u}, developed for image segmentation. U-Net's architecture has a symmetrical design, with contracting and expanding pathways, allowing for comprehensive feature extraction and precise localization of cancerous areas. U-Net's skip connections enable the fusion of low-level and high-level features, enhancing segmentation accuracy. This network's adaptability and proficiency in discerning fine details in cancer boundaries make it an ideal choice for segmenting cancerous regions within mpMRI. The network's innate ability to learn and adjust parameters caters to variations in mpMRI data, accommodating diverse patient anatomies and image qualities.   
\subsubsection{Classification}
\vspace{1em}
\textbf{\newline ResNet-50 DNN: } ResNet50 is a convolutional DNN first proposed by He \etal~\cite{resnet}. ResNet50 is a variant of the ResNet model that has 48 convolutional layers. This architecture uses residual learning to train ultra-DNNs by adding shortcut connections that skip three layers.
% The implementation employs the robust CNN ResNet-50 architecture for the classification of prostate cancer within mpMRI images. 
ResNet-50's deep layers and residual connections empower it to extract intricate spatial features from the complex image data, which is crucial for distinguishing between cancerous and healthy tissue. The model's hierarchical feature extraction, aided by residual blocks, contributes to its exceptional accuracy. Its ability to learn diverse features and patterns enables cancer classification. These facts encourage the implementation of ResNet-50 for the classification of PCa using mpMRI images. 
% offering a reliable tool for automated diagnosis and aiding clinical decision-making in mpMRI-based PCa assessment.
\newline
\textbf{Recurrent Neural Network: } 
In traditional neural networks, inputs and outputs are connected in a one-way manner, whereas RNNs ~\cite{rnn} iterate over elements of a sequence, performing the same operation for each component, with results depending on previous computations. Feedforward neural networks evolved into RNNs by incorporating feedback loops. RNN's sequential processing capability makes it particularly suited for capturing temporal dependencies in medical image data. RNNs can be promising architecture for leveraging the temporal characteristics of mpMRI images, contributing to more precise and reliable diagnoses. By analyzing sequential information, RNN aids in recognizing nuanced patterns and variations within the images, leading to improved PCa classification performance.

\begin{table*}[ht!]
\centering
\caption{\textbf{Patient-level classification and mpMRI PCa segmentation on the validation set.} The best findings are highlighted in bold, while dashes are used to indicate outcomes that haven't been reported in the literature.}
\resizebox{0.9\textwidth}{!}{
\begin{tabular}{||>{\cellcolor{yellow!10}}l||c c c c c c||}
\hline
         \textbf{\shortstack{Architectures}} & \textbf{\shortstack{Accuracy\\ (\%)}} & \textbf{\shortstack{F1\\ (\%)}} & \textbf{\shortstack{Precision\\ (\%)}} & \textbf{\shortstack{Recall\\ (Sens. (\%))}}  & \textbf{\shortstack{Spec.\\ (\%)}}  & \textbf{\shortstack{Dice}} \\
         \hline
         PCF-SEL-MR~\cite{mehta2021computer} & - & - & 63.0 & 75.0 & 55.0 &  -\\
         {FocalNet~\cite{cao2019joint}} & - & - & - & 89.7  & -& -\\
         Liu \etal~\cite{liu2009prostate} & 89.38 & - & - & 87.5 & 89.5 & 0.62 \\
         Artan \etal~\cite{artan2010prostate} & - & - & - & 85.0  & 50.0 & 0.34 \\
         Zhang \etal~\cite{zhang2021new} & 80.97 & - & 76.69 & -  & - & -\\
         
\hline
          DeepSegNet+ResNet50~\cite{deepsegnet} & 72.41 & 52.47 & 89.82 & 58.62 & 87.47  & 0.54\\
          
         DeepSegNet+RNN~\cite{deepsegnet} & 84.37 & 79.02 & 91.83 & 83.26 & 90.31 & 0.62\\
         
         U-Net+RNN & 92.29 & 88.43 & 92.04 & 91.92 & 89.89 & 0.65\\
         
         U-Net+LSTM (Ours) & \textbf{94.69} & \textbf{92.09} & \textbf{95.17} & \textbf{92.09}  & \textbf{96.88} & \textbf{0.67} \\
\hline
\end{tabular}}
\label{tab:1}
\end{table*}

 % DNN & 68.31 & 45.28 & 88.38 & 46.98 & 0.719 & 89.53 & 0.59\\
 %         Deep RNN & 86.43 & 88.37 & 88.43 & 89.53 & 0.787 & 91.81 & 0.64\\

\begin{figure*}[h!]%
    \centering
    \subfloat[\centering Accuracy values]{{\includegraphics[width=9cm]{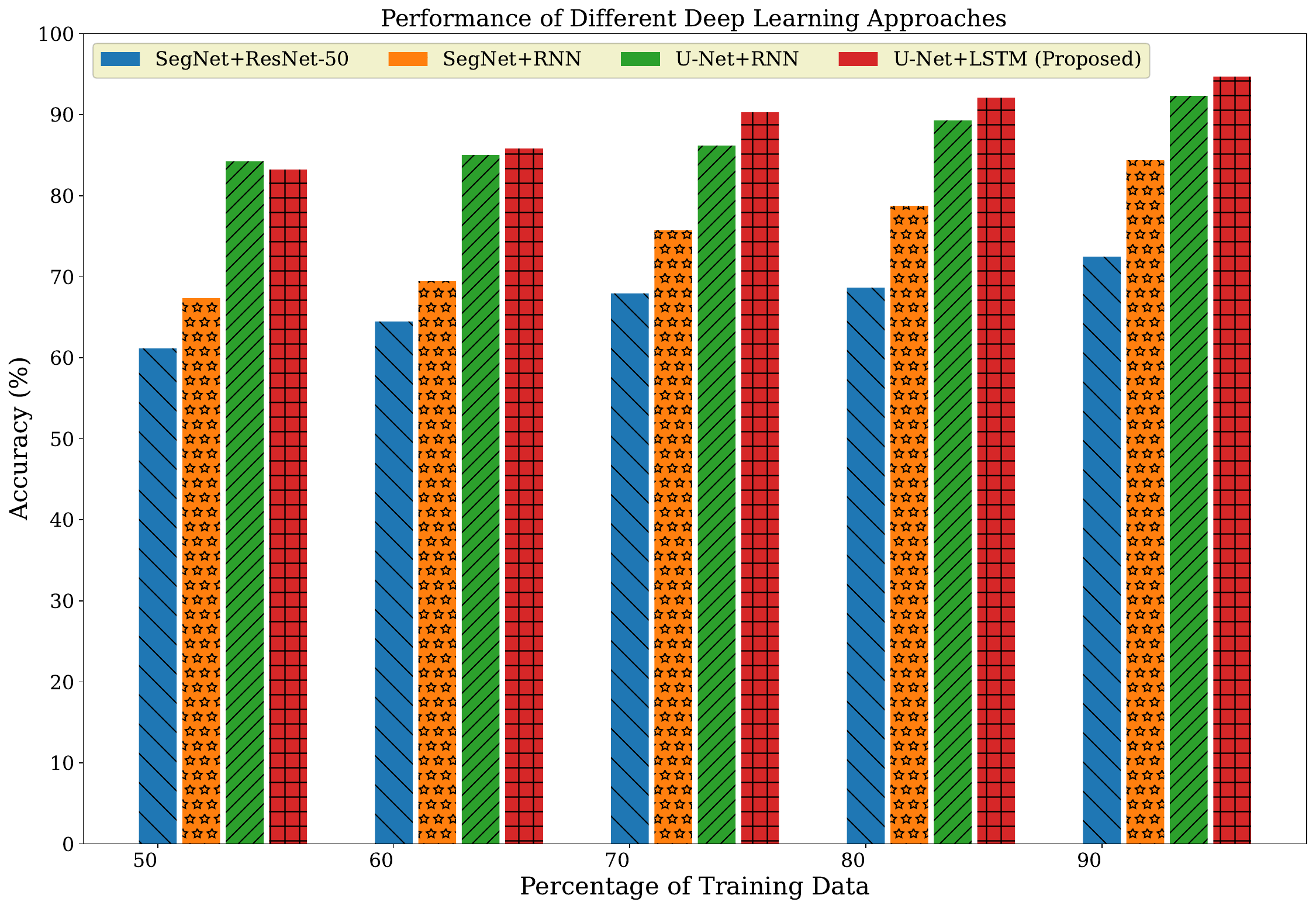} }}%
    \subfloat[\centering Sensitivity values]{{\includegraphics[width=9cm]{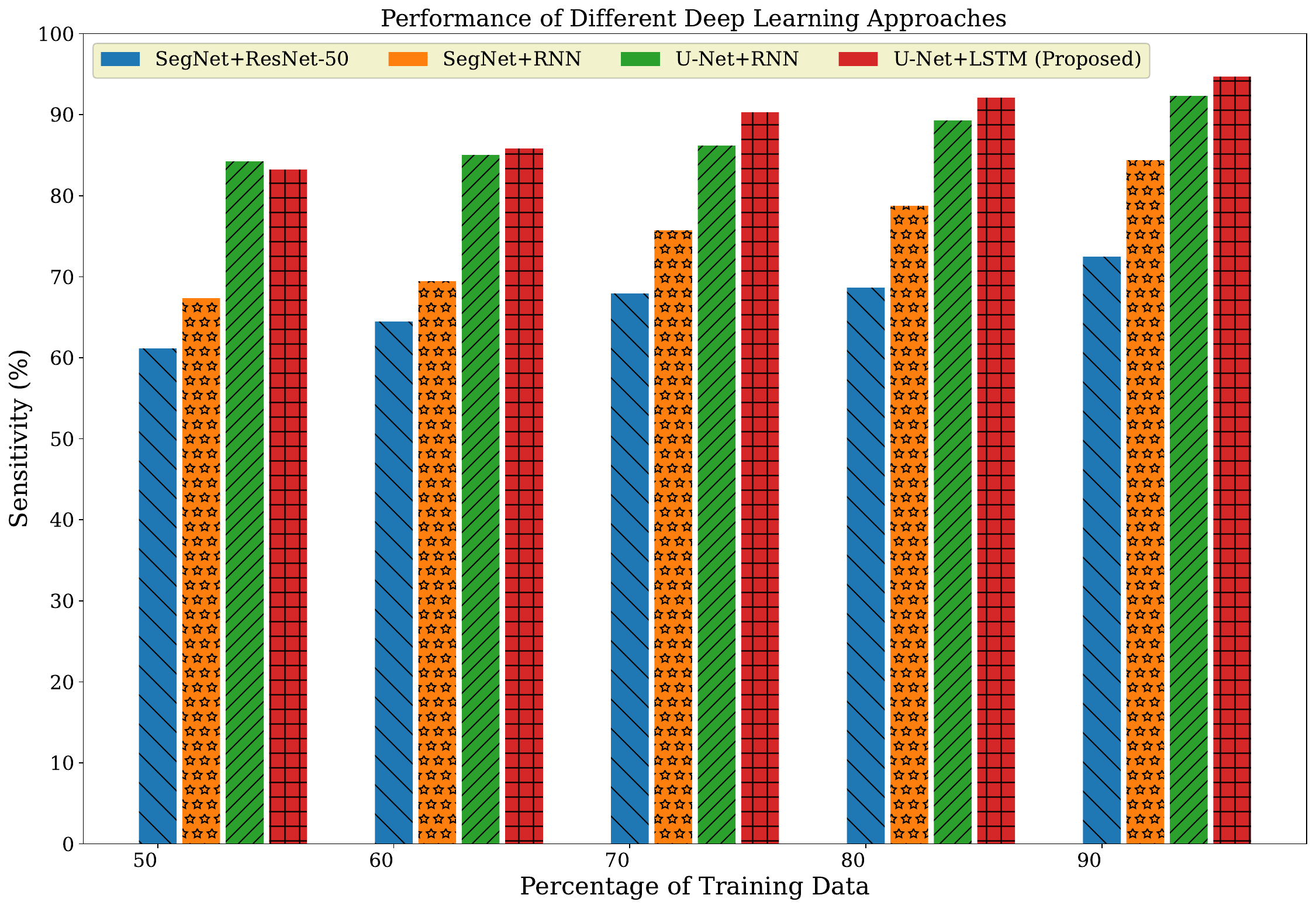} }}%
    \caption{\textbf{Evaluating four DL pipelines}. The x-axis shows the percentage of training data used in model development. The y-axis displays the values in percentage.}
    \label{fig2}
\end{figure*}

\textbf{Long Short-Term Memory: } RNN models were enhanced to address the vanishing gradient problem, which led to the development of the LSTM model~\cite{rnn}. A novel additive gradient structure that includes a forget gate has been introduced to address the vanishing gradient issue in LSTM.
By effectively accounting for temporal dependencies, the LSTM classifier aids in discerning intricate patterns and subtle variations, ultimately leading to more precise and robust cancer classification. This integration of LSTM underscores a promising avenue for exploiting mpMRI images, allowing them to remember or forget information as needed.
The utilization of an LSTM classifier forms a crucial component of the classifier part in our DL pipeline, aiming to improve classification results.

\subsection{Evaluation Metrics}
The confusion matrix (CM) is an important resource utilized in evaluating DL models.
CM is a square matrix that contains four categories true positives (TP), false positives (FP), false negatives (FN), and true negatives (TN). TP means the model identifies the positive class while the actual class is also positive, and vice versa for FN. In the FP case, the model incorrectly predicts the positive class even though the actual class is negative. Finally, TN cases arise when the model correctly predicts the negative class while the actual class is also negative.

Our DL pipeline is evaluated on validation results using accuracy, F1-score, Precision, Recall (also known as sensitivity), Specificity, and Dice Score. The accuracy metric assesses the model's overall precision, indicating the extent to which it produces correct results, and is calculated by $(TP+TN)/(TP+FP+FN+TN)$. Precision expresses the measurement of correctly identified positive instances among all instances predicted as positive and is determined by $TP/(TP+FP)$. Recall can be calculated using $TP/(TP+FN)$ and measures the accuracy of correctly predicting positive instances out of all the actual positive instances. F1 is the harmonic mean of precision and recall, calculated as $(2\times Recall\times Precision)/ (Recall+Precision)$. Specificity, calculated by $TN/TN+FP$, tells us how well the model can accurately predict that an observation does not belong to a specific category. Finally, the dice score is a metric used to evaluate the performance of image segmentation algorithms, and it is calculated as twice the intersection of the predicted and ground truth masks divided by the sum of their areas.

% \vspace{-2em}
\begin{figure*}[ht!]
    \centering
    \includegraphics[width=1\textwidth]{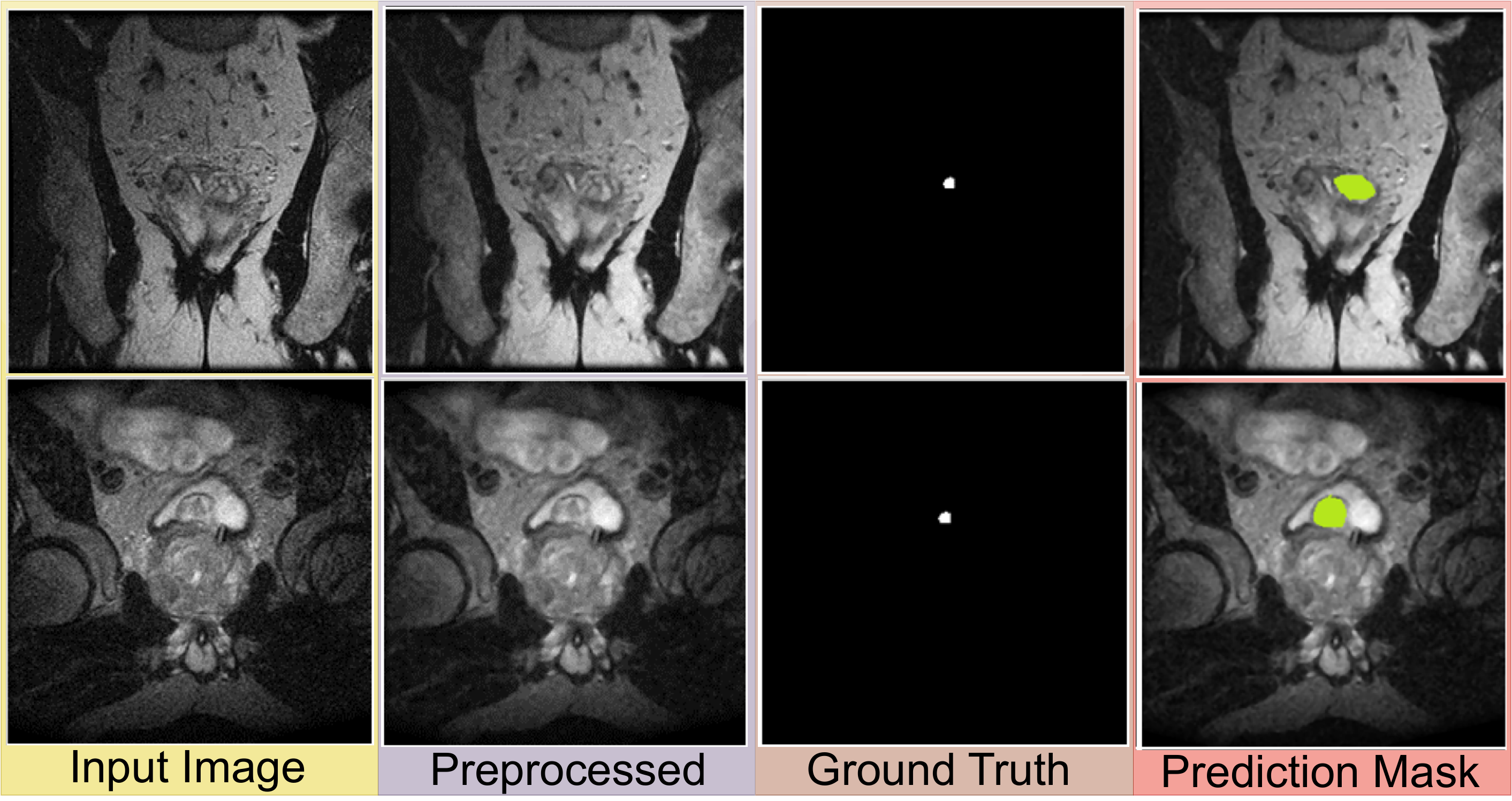}
    \caption{Visualization of the results from proposed DL pipeline. }
    \label{fig3}
\end{figure*}
% \vspace{-2em}

\subsection{Experimental Setup}
We used the Caffe framework to borrow the aforementioned DL models. The stochastic gradient descent (SGD) optimizer and a learning rate of 0.0003 were used for the hyperparametric search. Unpadded convolutions are used, and the resultant output images show a modest size reduction relative to the input image. The proposed DL pipeline compensates for each DL model's drawbacks so that each may be used to its full potential. \\

\section{Results and Discussion}\label{sec:discussion}

Table~\ref{tab:1} presents a comparative analysis of four DL architectures for mpMRI-based PCa segmentation and classification. Among our developed pipelines, those using U-Net outperform all other literature works in most evaluation metrics. The U-Net+LSTM model surpasses all others, significantly enhancing the accuracy and sensitivity of DeepSegNet+ResNet-50 by nearly 22\% and 33\%, respectively. In medical applications, sensitivity and specificity hold particular importance as they directly affect patient outcomes and safety, while accuracy provides a holistic view of model performance. In this regard, the second-best results come from U-Net+RNN, with a small difference in accuracy compared to U-Net+LSTM. In terms of segmentation, we achieved a small gain in finding accurate candidate regions compared to literature work but most of our developed pipeline gives a comparable number, the highest of which was 0.67, estimated by the U-Net+LSTM pipeline. These findings highlight the U-Net+LSTM DL pipeline`s potential to improve PCa diagnosis precisely using mpMRI images.

The impact of varying training percentages on different architectures is assessed using accuracy and sensitivity in Figure~\ref{fig2}(a) and Figure~\ref{fig2}(b). It can be observed that accuracy and sensitivity steadily improve as the training percentage increases from 50\% to 80\%. The models perform optimally between 80\% and 90\% training data, with minimal accuracy gains beyond 90\%. These insights guide architecture selection and data allocation for optimal model performance. In mpMRI-based PCa detection, overlay representations can efficiently highlight potential cancerous regions by integrating visual cues into the original images. In Figure~\ref{fig3}, we provide a comprehensive visualization of our best DL pipeline's segmentation outcomes. In Figure~\ref{fig3}(a), the initial input appears as a mpMRI, capturing the prostate gland's anatomical and tissue properties. Preprocessing in Figure~\ref{fig3}(b) standardizes image for DL models, while the ground truth in Figure~\ref{fig3}(c) denotes manually annotated cancerous regions. The prediction mask in Figure~\ref{fig3}(d) illustrates our U-net model's output, significantly delineating identified potential cancerous areas. This figure conveys the transformation of raw input into precise cancer ROI predictions, showcasing the effectiveness of our approach in localizing candidate regions for giving second opinions. 

\section{Conclusion and Future Work}\label{sec:conclusion}
Early identification is essential for lowering the mortality rate in prostate cancer (PCa) cases, which is the most common cancer among men worldwide. Using mpMRI imaging, deep learning (DL) can potentially increase the precision and efficacy of PCa diagnosis. DL approaches using mpMRI data have great potential in improving the accuracy of diagnosis and treatment planning for PCa. 
In this research work, we have proposed a DL-based method for identifying PCa using mpMRI, the U-Net model for image segmentation, and the LSTM model for final classification. For the segmentation problem, our suggested method attained a dice value of 0.67,   94.69\% accuracy, and of 92.09\% F1-score. The obtained results show that our strategy outperforms standalone techniques like ResNet50. The presented approach has considerable potential for increasing PCa diagnosis accuracy and efficacy and advancing the creation of clinical decision support systems that can automatically make PCa diagnoses. The proposed DL-based strategy has significant potential to automate future PCa diagnosis and therapy dramatically.

A single dataset has been used to develop and evaluate the proposed DL pipeline for PCa diagnosis using mpMRI, which limits its generalizability to other datasets. To enhance the generalizability of DL models, incremental learning may be used to tune models on other public datasets, such as PICTURE and PROSTATEx~\cite{prostate_2018}. To increase the precision of PCa grading, future studies can concentrate on integrating vision transformers (ViTs)~\cite{kanwal2023vision}. Perhaps in conjunction with data from other modalities,  mpMRI can potentially improve PCa diagnosis substantially using DL models.

\bibliographystyle{IEEEtran}
\bibliography{bibliography}% common bib file
\end{document}